\documentclass[aps,prb,twocolumn,showpacs,superscriptaddress]{revtex4}

\usepackage{graphicx}

\newcommand{\ef}{$E_F$}

\begin{document}

\title{Mott localization in a pure stripe antiferromagnet Rb$_{1-\delta}$Fe$_{1.5-\sigma}$S$_2$}
\author{Meng Wang}
\email{wangm@berkeley.edu}
\affiliation{Department of Physics, University of California, Berkeley, CA 94720, USA }
\author{Ming Yi}
\affiliation{Department of Physics, University of California, Berkeley, CA 94720, USA }
\author{Huibo Cao}
\affiliation{Quantum Condensend Matter Division, Oak Ridge National Laboratory, Oak Ridge, TN 37831, USA}
\author{C. de la Cruz}
\affiliation{Quantum Condensend Matter Division, Oak Ridge National Laboratory, Oak Ridge, TN 37831, USA}
\author{S. K. Mo}
\affiliation{Advanced Light Source, Lawrence Berkeley National Lab, Berkeley, CA 94720, USA}
\author{Q. Z. Huang}
\affiliation{NIST Center for Neutron Research, National Institute of Standards and Technology, Gaithersburg, MD 20899, USA}
\author{E. Bourret-Courchesne}
\affiliation{Materials Science Division, Lawrence Berkeley National Laboratory, Berkeley, CA 94720, USA }
\author{Pengcheng Dai}
\affiliation{Department of Physics and Astronomy, Rice University, Houston, TX 77005, USA}
\author{D. H. Lee}
\affiliation{Department of Physics, University of California, Berkeley, CA 94720, USA }
\affiliation{Materials Science Division, Lawrence Berkeley National Laboratory, Berkeley, CA 94720, USA }
\author{Z. X. Shen}
\affiliation{Stanford Institute for Materials and Energy Sciences, SLAC National Accelerator Laboratory and Stanford University, Menlo Park, CA 94025, USA }
\affiliation{Department of Physics and Applied Physics, and Geballe Laboratory for Advanced Materials, Stanford University, Stanford, CA 94305, USA }
\author{R. J. Birgeneau}
\affiliation{Department of Physics, University of California, Berkeley, CA 94720, USA }
\affiliation{Materials Science Division, Lawrence Berkeley National Laboratory, Berkeley, CA 94720, USA }
\affiliation{Department of Materials Science and Engineering, University of California, Berkeley, CA 94720, USA }

\begin{abstract}

A combination of neutron diffraction and angle-resolved photoemission spectroscopy measurements on a pure antiferromagnetic stripe Rb$_{1-\delta}$Fe$_{1.5-\sigma}$S$_2$ is reported. A neutron diffraction experiment on a powder sample shows that a 98$\%$ volume fraction of the sample is in the antiferromagnetic stripe phase with rhombic iron vacancy order and a refined composition of Rb$_{0.66}$Fe$_{1.36}$S$_{2}$, and that only 2$\%$ of the sample is in the block antiferromagnetic phase with $\sqrt{5}\times \sqrt{5}$ iron vacancy order. Furthermore, a neutron diffraction experiment on a single crystal shows that there is only a single phase with the stripe antiferromagnetic order with the refined composition of Rb$_{0.78}$Fe$_{1.35}$S$_2$, while the phase with  block antiferromagnetic order is absent. Angle-resolved photoemission spectroscopy measurements on the same crystal with the pure stripe phase reveal that the electronic structure is gapped at the Fermi level with a gap larger than 0.325 eV. The data collectively demonstrate that the extra 10$\%$ iron vacancies in addition to the rhombic iron vacancy order effectively impede the formation of the block antiferromagnetic phase; the data also suggest that the stripe antiferromagnetic phase with rhombic iron vacancy order is a Mott insulator. 

\end{abstract}

\maketitle




The discovery of superconductivity in materials with the composition $A_{x}$Fe$_{2-y}$Se$_2$ ($A=$ K, Tl/K) at temperatures above 30 K has stimulated intense interest in the iron chalcogenides\cite{Hsu2008, Yeh2008, Guo2010, Fang2010}. Since their original discovery, superconductivity has been observed in a large number of metallic\cite{Wangaf2011, Krzton2011,Ying2012} and molecular spacer layer intercalated\cite{Burrard2012,Lu2014b} FeSe-based materials, as well as FeSe monolayers on a SrTiO$_3$ substrate\cite{Wang2012}. The $A_{x}$Fe$_{2-y}$Se$_2$ system is fascinating because the superconducting (SC) compounds lack the hole Fermi pockets at the Brillouin zone center that have been deemed necessary for the proposed antiferromagnetic fluctuation-driven pairing based on the Fermi surface nesting picture developed for the Fe pnictide superconductors\cite{Mazin2008,Zhang2011a,Mou2011,Qian2011}; nevertheless, the Fe chalcogenides superconduct at temperatures comparable to those in the other iron-based superconductors. Furthermore, the chalcogenide compounds exhibit a number of interesting structural and magnetic phases. Specifically, the SC compounds studied to-date always consist of a SC phase and an insulating block antiferromagnetic (AF) phase with $\sqrt{5}\times \sqrt{5}$ iron vacancy order. This insulating block AF phase, which is commonly observed, has a composition close to that of the stoichiometric $A_{0.8}$Fe$_{1.6}$Se$_2$ compound (referred to as the ``245" phase) and exhibits large moments ($\sim 3.3\mu_B$) aligned along the $c$ axis in the AF state\cite{Bao2011, Chen2011a, Dagotto2013}. 

\begin{figure}[t]
\includegraphics[scale=0.4]{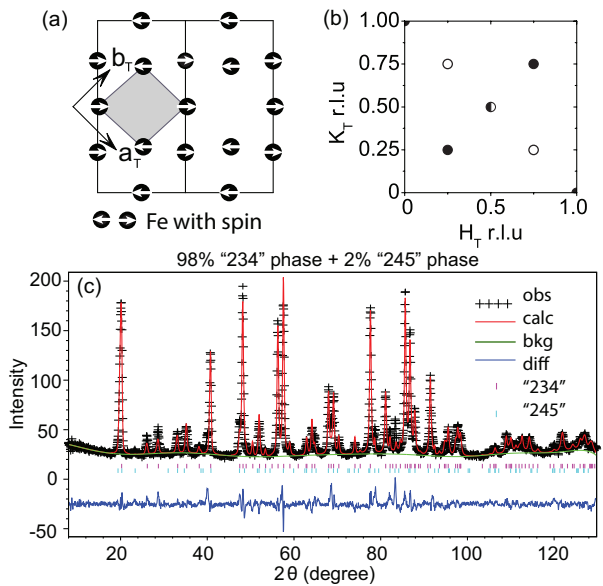}
\caption{(Color online).  (a) A schematic for the iron plane with rhombic vacancy order of the 234 phase. The grey shaded area represents the tetragonal unit cell we used. (b) The reciprocal space corresponding to the tetragonal unit cell, which is the notation used throughout the paper.  The closed and open circles show the wave vectors of the magnetic peaks at $L=odd$ and the nuclear peaks at $L=even$ of the 234 phase in the $[H_T, K_T]$ plane, respectively. (c) Neutron powder diffraction spectrum at 500 K. Refinement shows a $98\%$ volume fraction in the 234 phase with a composition of Rb$_{0.66}$Fe$_{1.36}$S$_2$ and a $2\%$ volume fraction in the 245 phase with a composition of Rb$_{0.8}$Fe$_{1.6}$S$_2$.  }
\label{fig1}
\end{figure}

In compounds where the iron content is less than 1.6, in addition to the 245 phase, a second phase with a stripe AF order and rhombic iron vacancy order, as presented in Fig.~\ref{fig1}(a), is also found\cite{Wang2011a, Zhao2012, Wangm2014}. This stripe AF phase, as discovered in TlFe$_x$S$_2$, K$_{0.81}$Fe$_{1.58}$Se$_2$, and Rb$_{0.8}$Fe$_{1.5}$S$_2$ has a composition close to stoichiometric $A$Fe$_{1.5}X_2$ ($X$ = Se, S), and hence is referred to as the ``234" phase\cite{Sabrowsky1986,Zhao2012, Wangm2014}. Interestingly, the 234 phase in K$_{0.81}$Fe$_{1.58}$Se$_2$ and Rb$_{0.8}$Fe$_{1.5}$S$_2$ have the same moment size of $2.8\mu_B$ and very similar N$\acute{\mathrm{e}}$el temperatures of 280 K and 275 K, respectively. Inelastic neutron scattering  studies on the spin waves of the stripe AF order suggest that the moments are fully localized and are in a high spin configuration $S=2$, in clear contrast to the presence of itinerant magnetism in the iron pnictides\cite{Zhao2014, Wang2015}. However, thermal activation gaps fitted from resistivities on the 234 and 245 interdigitating samples and first-principle calculations on the 234 phase yield a small electronic gap ($\sim 40$ meV) to the Fermi level, inconsistent with fully localized magnetism\cite{Zhao2012, Wangm2014,Yan2011}. Most interestingly, the stripe AF phase has the same magnetic structure as the parent compounds of the iron pnictide superconductors, and hence has been proposed as a possible parent compound of the superconductors in the iron chalcogenides. However, the ubiquitous coexistence of the robust 245 phase has thus far prevented a precise determination of the intrinsic properties of the 234 stripe phase.

As has been realized previously in studies of the phase diagrams of the La$_2$CuO$_{4+y}$ superconductors\cite{Wells1997}, in order to understand the behavior of the Fe pnictide and chalcogenide superconductors, it will be necessary to characterize all of the nearby phases, which are realized by changes in the doping and stoichometry.

In this paper, we report neutron diffraction and angle-resolved photoemission spectroscopy (ARPES) studies on nominal Rb$_{0.8}$Fe$_{1.5}$S$_2$ samples. Refinement from powder and single crystal samples yields compositions of Rb$_{0.66}$Fe$_{1.36}$S$_2$ and Rb$_{0.78}$Fe$_{1.35}$S$_2$, respectively, and, importantly, shows the absence of the 245 phase in the single crystal. This demonstrates that the existence of 10$\%$ additional randomly distributed iron vacancies effectively impedes the formation of the 245 phase in the iron chalcogenides.  Said in a different way, reducing the Fe concentration below 1.5 means that one no longer has two phase coexistence between the 1.5 Fe stripe phase and the 1.6 Fe block phase. Furthermore, ARPES measurements on the sample with only the pure 234 phase reveal a gap as large as 0.325 eV at the Fermi level, which is consistent with theoretical predictions that suggest that the 234 phase is a Mott localized system promoted by the iron vacancies\cite{Yu2011, Cao2011}.

\begin{table}[t]
\centering
\caption{Structure parameters of Rb$_{0.66}$Fe$_{1.36}$S$_2$ at 500 K. Space group $Ibam$ (No. 72) with $a_o=5.5235(3)$, $b_o=10.7847(7)$, and $c=13.944(1)$ \AA . $Rp=6.11\%$, $wRp=8.91\%$, $\chi^2=0.732$.}
\label{my-label}
\begin{tabular}{@{}*{6}{p{.08\textwidth}@{}}}
\hline \hline
Atom & Site & x     & y     & z     & Occup. \\ \hline
Rb   & 8i   & 0.240(2) & 0.117(1) & 0     & 0.66(2)  \\
Fe   & 8j   & 0     & 0.2432(4) & 0.25  & 0.90(2)  \\
Fe   & 4b   & 0.5   & 0     & 0.25  & 0.93(2)   \\
S    & 16k  & 0.228(2) & 0.106(1) & 0.3352(7) & 1.00      \\ \hline \hline
\end{tabular}
\label{table:t1}
\end{table}

\begin{figure}[t]
\includegraphics[scale=0.6]{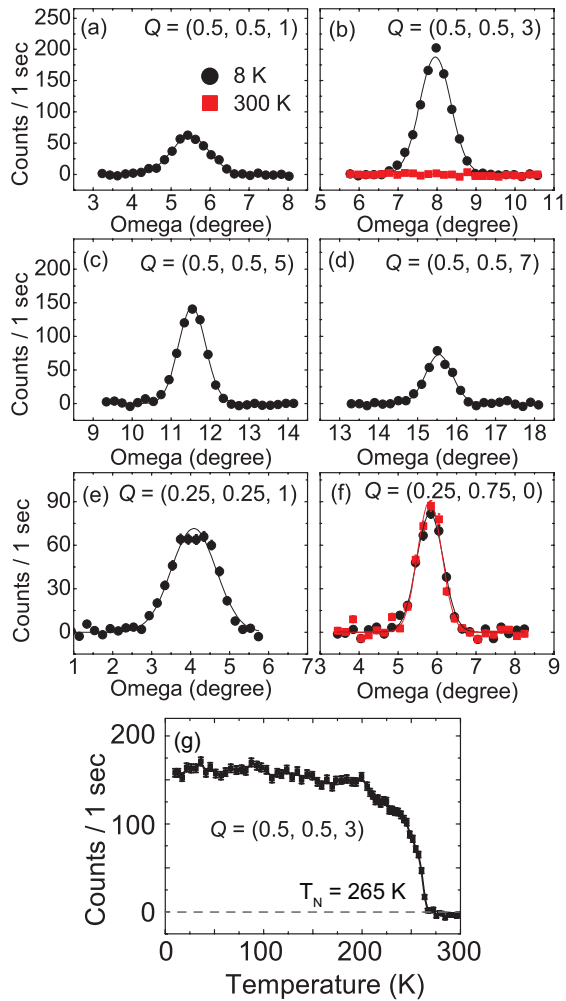}
\caption{(Color online).  (a-d) Rocking curve scans of the magnetic peaks associated with the 234 phase through $Q = (0.5, 0.5, L)$ with $L = 1, 3, 5$, and 7. (e, f) Scans through the fingerprint reflection peaks of the rhombic iron vacancy order at $Q = (0.25, 0.25, 1)$ (magnetic) and $Q=(0.25, 0.75,0)$ (nuclear). The data are collected at 8 K (black) and 300 K (red), respectively. (g) Temperature dependence of the magnetic peak at $Q = (0.5, 0.5, 3)$ reveals the N$\acute{\mathrm{e}}$el temperature of $T_N = 265$ K. }
\label{fig2}
\end{figure}

The single crystals were grown using the Bridgman method\cite{Wangm2014}. Neutron diffraction experiments were carried out on both the HB-2A powder diffractometer and the HB-3A four-circle single crystal diffractometer at the High-Flux Isotope Reactor, Oak Ridge National laboratory. For the powder diffraction experiment at HB-2A, 2 g of single crystals were ground into powder. Neutrons with wavelength $\lambda=2.41$ \AA\ and a cryofurnace that can reach up to 800 K were employed. For the single crystal diffraction experiment at HB-3A, the wavelength of $\lambda=1.003$ \AA\ was used from a bent perfect Si-331. A single piece of crystal weighing 58 mg with a mosaic of 1$^\circ$ full width at half maximum was loaded into a closed-cycle refrigerator.  ARPES experiments were performed at beamline 10.0.1 of the Advanced Light Source, Lawrence Berkeley National Laboratory. An R4000 electron analyzer was used to allow an energy resolution better than 15 meV and angular resolution of 0.3$^\circ$. The samples were cleaved in situ and measured in ultra-high vacuum with a base pressure better than 4$\times$10$^{-11}$ torr. All data were measured by 45 eV photons. The ARPES measurements for the insulating compound were done at 130 K to avoid charging effects. A photon flux-dependence test was performed to ensure that no noticeable shift energy was observable at the flux used.

Previous studies on nominal Rb$_{0.8}$Fe$_{1.5}$S$_2$ single crystals revealed a mesoscopic interdigitation of the 234 phase and the 245 phase with N$\acute{\mathrm{e}}$el temperatures of $T_N^{234}=275$ K and $T_N^{245}=425$ K, respectively\cite{Wangm2014}.  In Fig.~\ref{fig1}(c), we present a neutron powder diffraction spectrum on the 2 g powder sample at 500K, which is above both magnetic transitions in order to avoid magnetic reflections. Surprisingly, refinement on the powder spectrum shows that a 98$\%$ volume fraction of the sample is in the 234 phase and only a 2$\%$ volume fraction is in the commonly observed 245 phase. The lattice constants of the 245 phase are refined to be $a=b=8.537(2)$\AA , and $c=13.928(5)$\AA\  in the tetragonal structure, $I4/mmm$. We list the refined parameters of the 234 phase in Table \ref{table:t1}. The refined composition of Rb$_{0.66}$Fe$_{1.36}$S$_2$ reveals that there are approximately 10$\%$ iron vacancies randomly distributed among the iron sites in addition to the ordered rhombic iron vacancies. The high volume fraction of the 234 phase suggests that it is possible to obtain a pure 234 phase in a single crystal.

\begin{figure}[t]
\includegraphics[scale=0.43]{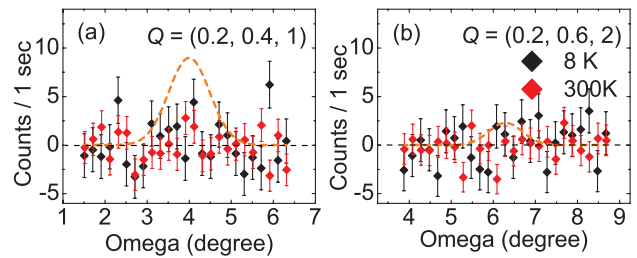}
\caption{(Color online).  (a) Rocking curve scans through the fingerprint reflection wave vectors of the block AF order at $Q = (0.2, 0.4, 1)$ (magnetic) (b) the $\sqrt{5}\times\sqrt{5}$ iron vacancy order at $Q = (0.2, 0.6, 2)$ (nuclear) at 8 K (black) and 300 K (red), respectively. The dashed orange lines are simulations assuming the existence of 1$\%$ of the 245 phase. }
\label{fig3}
\end{figure}

Next, we present elastic neutron scattering measurements on an irregularly shaped single crystal weighing 58 mg. The data at the wave vectors associated with the stripe AF phase are presented in Fig.~\ref{fig2}. To compare these results directly with previous reports\cite{Zhao2012, Wangm2014}, we employ the tetragonal notation with an averaged $a_T=b_T=3.879$ \AA\ and $c=13.773$ \AA (whereas $a_o=5.582, b_o=10.780$ and $c=13.773$ \AA\ in the orthorhombic $Ibam$ structure optimized at 8 K). We label the wave vectors as $Q=(H_T, K_T, L)$, where the $(H_T, K_T, L)$ are the Miller indices in the momentum transfer of $(q_x, q_y, q_z)=(2\pi H_T/a_T, 2\pi K_T/b_T, 2\pi L/c)$ in reciprocal lattice units (r.l.u). Figs.~\ref{fig2}(a)-\ref{fig2}(d) show rocking curve scans through the stripe magnetic peaks at $Q=(0.5, 0.5, L)$ with $L=1, 3, 5$, and 7 at 8 K. The intensities of the magnetic peaks at different $L$ follow the Fe$^{2+}$ magnetic form factor and are consistent with the in-plane ordered stripe AF order. A scan through $Q=(0.5, 0.5, 3)$ at 300 K in Fig.~\ref{fig2}(b) is flat, demonstrating that the temperature is above the N$\acute{\mathrm{e}}$el temperature of the 234 phase. The peaks as shown in Fig.~\ref{fig2}(e) at $Q=(0.25, 0.25, 1)$ and Fig.~\ref{fig2}(f) at $Q=(0.25, 0.75, 0)$ correspond to magnetic and nuclear reflections, respectively, induced by the presence of the rhombic iron vacancy order\cite{Wangm2014}. The temperature dependence of the peak intensities at $Q=(0.5, 0.5, 3)$ reveals that the N$\acute{\mathrm{e}}$el temperature is 265 K, slightly lower than that of the sample studied in our previous report\cite{Wangm2014}.

\begin{figure}[t]
\includegraphics[scale=0.45]{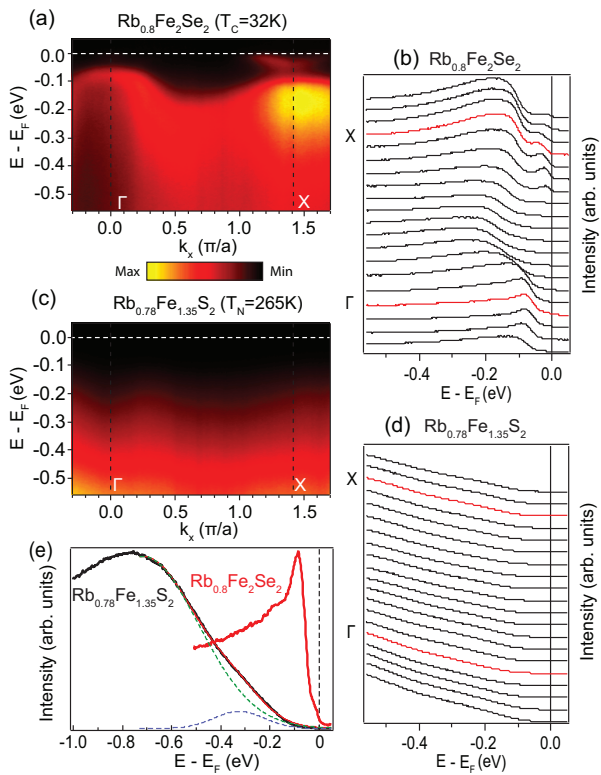}
\caption{(Color online).  (a) ARPES measurements of the electronic structures along the $\Gamma-X$ high symmetry cut on SC Rb$_{0.8}$Fe$_2$Se$_2$ ($T_c = 32$ K) at 30 K and (c) insulating Rb$_{0.78}$Fe$_{1.35}$S$_2$ ($T_N = 265$ K) at 130 K. The intensities are plotted in a linear scale.Their respective EDCs are shown in (b, d). (e) A comparison of the EDCs at $\Gamma$ point for Rb$_{0.8}$Fe$_2$Se$_2$ (red) and Rb$_{0.78}$Fe$_{1.35}$S$_2$ (black). The EDC for insulating Rb$_{0.78}$Fe$_{1.35}$S$_2$ is fitted with a half Gaussian background (green) and a Gaussian peak (blue), revealing a weak feature at -0.325 eV, in contrast to the sharp peak near the Fermi level for superconducting Rb$_{0.8}$Fe$_2$Se$_2$. }
\label{fig4}
\end{figure}

We present measurements on the 245 phase in Fig.~\ref{fig3}. Interestingly, scans through the magnetic wave vector $Q=(0.2, 0.4, 1)$ of the 245 phase and the fingerprint reflection peak at $Q=(0.2, 0.6, 2)$ of the  $\sqrt{5}\times \sqrt{5}$ iron vacancy order at both 8 K and 300 K are flat. We plot simulations assuming the existence of 1$\%$ 245 phase\cite{Wangm2014}. The magnetic peak intensities at $Q=(0.2, 0.4, 1)$ are clearly above the error bars as shown in Fig.~\ref{fig3}(a). Thus, our results demonstrat that the 245 phase is absent in this irregularly shaped single crystal.  Furthermore, refinement based on rocking curves through all the reachable nuclear peaks reveals that the refined composition of the sample is Rb$_{0.78}$Fe$_{1.35}$S$_2$. The iron contents as refined from the powder and single crystal samples are consistent, yielding the existence of $10\%$ extra iron vacancies in addition to the rhombic iron vacancy order. This puts the Fe compositions on average outside of the 1.5 to 1.6 two phase coexistence region for the 234 and 245 phases.  The rubidium content shows some deviation between the two refinements.

Knowing that the single crystal studied with neutron diffraction is of the pure 234 phase, we used ARPES to measure the electronic structures on a piece of single crystal that was cleaved from the same piece of Rb$_{0.78}$Fe$_{1.35}$S$_2$ used for the neutron diffraction experiment. We then compared the results to those from a superconducting crystal of Rb$_{0.8}$Fe$_{2}$Se$_2$  with $T_c = 32$ K. The SC Rb$_{0.8}$Fe$_{2}$Se$_2$ exhibits clearly dispersive bands near the Fermi level, resulting in electron Fermi surfaces near the $X$ point, as shown in Figs.~\ref{fig4}(a) and \ref{fig4}(b)\cite{Yi2013}. In contrast, measurements under the same conditions on our crystal Rb$_{0.78}$Fe$_{1.35}$S$_2$ show the absence of dispersive bands near the Fermi level, \ef, as shown in Figs.~\ref{fig4}(c) and \ref{fig4}(d). Identical measurements on the SC Rb$_{0.8}$Fe$_{2}$Se$_2$ and related metallic Rb$_{0.8}$Fe$_{2}$S$_2$ show bands that disperse in the $k_z$ direction, indicating that these measurements reveal bulk-like properties\cite{Yi2015}. The energy distribution curves (EDCs) at the $\Gamma$ point for  Rb$_{0.8}$Fe$_{2}$Se$_2$ and Rb$_{0.78}$Fe$_{1.35}$S$_2$ are compared in Fig.~\ref{fig4}(e), where the purely 234 sample has suppressed density of states near \ef, in contrast to the SC compound which shows a prominent peak near \ef. The most prominent feature in the 234 sample is a non-dispersive peak near -0.8 eV. A fit using a Gaussian as the background reveals a weak feature at $-0.325$ eV in Rb$_{0.78}$Fe$_{1.35}$S$_2$, indicating that the energy gap across \ef~for this compound is at least as large as 0.325 eV. We note that both the -0.8 eV peak and the -0.325 eV hump do not show noticeable dispersion in momentum. As shown by theoretical calculations\cite{Cao2011}, the SDW ordered ground state in the non-correlated limit ($U = 0$) for the 234 phase has a vanishingly small gap at $E_F$. Only with strong electron correlations ($U > 5$ eV) can a gap as large as 0.325 eV appear. Hence the origin of this large gap is Mott localization, in sharp contrast to the metallic SDW state of the parent iron pnictides. The lack of dispersion of the -0.8 eV peak and the -0.325 eV hump is consistent with this localized nature. We note that insulating behavior seen in the suppression of density of states at \ef~has been reported previously by ARPES studies in the insulating $A_{x}$Fe$_{2-y}$Se$_2$ compounds\cite{Chen2011a,Yi2013}. However, in those cases, the insulating samples always had the ubiquitous 245 phase present, and hence it is difficult to determine the origin of the insulating behavior. Here, the sample we have measured has a single pure 234 phase, and hence our observation of the large gap must be an intrinsic property of this stripe magnetic phase alone.

The large gap revealed by ARPES measurements is consistent with the fact that the Fe$^{2+}$ ions of the 234 phase have fully localized moments and a spin configuration of $S=2$ as derived from our inelastic neutron scattering experiment\cite{Wang2015}, and thus indicates that the 234 phase is a Mott insulator with one electron on each of the four Fe$^{2+}$ 3$d$  orbitals ($d_{xy}, d_{xz/yz}$, and $d_{x^2-y^2}$). The reduced thermal activation gaps fitted from resistivity curves could be due to the internal pressure on interfaces enhanced by phase separation and twinning effects, as it has been shown that the external pressure could decrease the thermal activation gaps in the $A_x$Fe$_y$Se$_2$ system\cite{Guo2012}. Alternatively, it could also be from a very small undetectable quasiparticle spectral weight of the 234 phase. Further research using other probes such as optical measurements is necessary to determine the origin of this discrepancy.  

The discovery of a Mott insulating phase in the iron chalcogenides is reminiscent of the parent compounds of the cuprates, where one can progressively tune the Mott insulating compound to superconductivity by carrier doping\cite{Kastner1998}. However, in all likelihood the Mott localization and the stripe AF order of the 234 phase studied here are direct results of a particular rhombic iron vacancy order, which inhibits the itineracy of the electrons. This is similar in nature to the Mott insulating block AF order stabilized in the $\sqrt{5}\times \sqrt{5}$ iron vacancy ordered  245 phase, as well as the Mott localization realized in the Fe-diluted NaFe$_{0.5}$Cu$_{0.5}$As\cite{Song2015}. Incidentally, it has not been shown that the 234 phase with Fe content close to 1.5 can be tuned continuously to the superconducting phase, where the Fe content is close to 2\cite{Texier2012,Shoemaker2012,Carr2014,Kobayashi2015}. Rather as the Fe content is increased there is a first order transition from the 234 phase to the 245 phase where the Fe content is approximately 1.6 followed by a second first order transition to the metallic phase with Fe content close to 2\cite{Wangm2014}. Thus, the relationship between the stripe AF 234 phase and the superconducting phase in the iron chalcogenides appears to be different than those of the metallic collinear AF ordered parent compounds of the iron pnictides and the Mott insulating checkboard AF ordered parent compounds of the cuprates and their perspective doped superconducting variants.

In summary, we have successfully obtained a pure stripe AF phase with rhombic iron vacancy order. Our ARPES measurements show an insulator with a band gap as large as 0.325 eV. Together with previous inelastic neutron scattering measurements showing a spin configuration of $S=2$, our results reveal that the 234 phase is a Mott insulator. The refined composition suggests that the 10$\%$ randomly distributed iron vacancies in addition to the rhombic iron vacancy order enhance the localization and exclude the block AF ordered 245 phase. Further measurements probing the gap behavior in Rb$_{0.78}$Fe$_{1.35}$S$_2$ using optical and resonant inelastic x-ray techniques would be invaluable.

This work is supported by the Director, Office of Science, Office of Basic Energy Sciences, US Department of Energy, under Contract No. DE-AC02-05CH11231 and DE-AC03-76SF008. The research at Oak Ridge National Laboratory's High-Flux Isotope Reactor and Lawrence Berkeley National Laboratory's Advanced Light Source are sponsored by the Scientific User Facilities Division, Office of Basic Energy Sciences, US Department of Energy. Work at Rice is supported by the US DOE, BES under contract no. DE-SC0012311 (P.D.). Work at Stanford is supported by the DOE Office of Basic Energy Sciences, Division of Materials Sciences.



\bibliography{mengbib}

\end{document}